\documentclass[aps,pra,reprint,floatfix,superscriptaddress]{revtex4-1}
\usepackage{color}   
\usepackage{graphicx}
\usepackage{amsmath}
\usepackage{amsfonts}
\usepackage[T1]{fontenc}

\begin{document}
\renewcommand*{\figurename}{FIG. }
\newtheorem{thm}{Theorem}
\newtheorem{prop}{Proposition}
\newenvironment{proof}[1][\textit{Proof}]{\begin{trivlist}
\item[\hskip \labelsep {#1}]}{\end{trivlist}}

\def\be{\begin{equation}}
\def\ee{\end{equation}}
\def\bea{\begin{eqnarray}}
\def\eea{\end{eqnarray}}
\def\bi{\begin{itemize}}
\def\ei{\end{itemize}}
\def\bin{\begin{enumerate}}
\def\ein{\end{enumerate}}
\def\la{\langle}
\def\ra{\rangle}
\newcommand{\mv}[1]{\left\la#1\right\ra}
\newcommand{\vect}[1]{\mathbf{#1}}
\newcommand{\tothink}[1]{\textcolor{blue}{#1}}
\newcommand{\toadd}[1]{\textcolor{green}{#1}}
\newcommand{\todel}[1]{\textcolor{red}{#1}}

\newcommand{\new}[1]{\textcolor{cyan}{#1}}
\newcommand{\tmp}[1]{\textcolor{magenta}{#1}}

\title{One-dimensional $s$-$p$ optical superlattice}

\author{Wojciech Ganczarek}
\affiliation{Marian Smoluchowski Institute of Physics, Jagiellonian University, ul.\
Reymonta 4, 30-059 Krak\'ow, Poland}

\author{Michele Modugno}
\affiliation{Department of Theoretical Physics and History of Science, UPV-EHU, 48080 Bilbao, Spain}
\affiliation{IKERBASQUE, Basque Foundation for Science, 48011 Bilbao, Spain}

\author{Giulio Pettini}
\affiliation{Dipartimento di Fisica e Astronomia, Universit\`a di Firenze,
and INFN, 50019 Sesto Fiorentino, Italy}

\author{Jakub Zakrzewski}
\affiliation{Marian Smoluchowski Institute of Physics, Jagiellonian University, ul.\
Reymonta 4, 30-059 Krak\'ow, Poland}
\affiliation{Mark Kac Complex Systems Research Center, Jagiellonian University, ul.\ 
Reymonta 4, 30-059 Krak\'ow, Poland}


\begin{abstract}
The physics of one dimensional optical superlattices with resonant $s$-$p$ orbitals
is reexamined in the language of appropriate Wannier functions. 
It is shown that details of the tight binding model
realized in different optical potentials crucially depend on the proper determination
of Wannier functions. We discuss the properties of a 
superlattice model which quasi resonantly couples $s$ and $p$ orbitals and show its
relation with different tight binding models used in other works. 
\end{abstract}


\maketitle

\section{Introduction}
Cold atoms in optical lattices form a versatile medium for realizing different models
of many-body systems, ranging from condensed matter to high-energy physics 
\cite{Jaksch05,Lewenstein07,hep}. While originally the attention has been restricted
to the standard simplest models (such as the celebrated Bose-Hubbard
 tight-binding Hamiltonian \cite{fisher89,jaksch98, Greiner02}), soon there appeared
 a vast proliferation of research exploring
unique possibilities offered by internal atomic structure (e.g. offering the possibility
to create artificial gauge fields 
\cite{Lewenstein12,Taglia12,Taglia13}) or by the flexibility of optical lattice potentials. 
In particular, superposition of 
laser standing waves with different wavevectors allows to create superlattice (SL) potentials.
When these wavevectors are 
incommensurable, the resulting optical potential is pseudo-random, enabling studies of 
disordered systems \cite{Damski03,Fallani07,Fallani08,Zakrzewski09}. 
Small wavevector integer ratios lead to double- or triple-well periodic potentials,
extensively studied due to their interesting properties
 \cite{Trotzky08,hemmerich}. In particular, even simple one-dimensional potentials
 allow to observe interesting topological properties as the 
Zak-Berry phase \cite{Bloch13} or topological edge states \cite{Grusdt13}.
Depending on the details of the model studied, 
different novel phases have been predicted \cite{Dhar2011}. 

Most of these interesting studies assume a mapping between a continuous and a discrete version 
of the Hamiltonian, mapping accomplished by an appropriate 
choice of Wannier functions. Quite frequently, a discrete version of the Hamiltonian has been 
studied for arbitrary values of its parameters 
\cite{Dhar2011,dhar13}. The latter are, however, typically uniquely defined by the
nature of the optical potentials and, possibly, interactions. 
Their determination requires a proper choice of the discrete basis representing the lattice.
While for one dimensional sinusoidal potentials the procedure
 of constructing such a basis is well known from Kohn seminal works \cite{kohn}, only recently
 a method to obtain the optimal basis of maximally localized Wannier
functions for a double-well SL potential has been developed \cite{Modugno}. This approach
relies on the general scenario of Marzari and Vanderbilt \cite{Marzari97, Marzari12} 
and consists in designing  a specific two-step gauge transformation of the Bloch functions 
for a composite two-band system. 

Often
one is interested in SL potentials which enable efficient coupling between the ground and excited bands, 
as exemplified in experiments of Hemmerich group
 \cite{hemmerich}. The orbital physics \cite{LiuWu06,Wu06,Wu07,Wu08,Wu09} in such SL potentials \cite{Cai11,Li13,Li13a,p-realization,soltan} 
 lead to novel physical situations. As the method developed to
 find the optimal Wannier basis \cite{Modugno} is valid for a generic set of two bands we apply
 it in this work considering in detail a SL case with a resonance 
between $s$ and $p$ orbitals in the neighboring sites.  Besides, in Section \ref{resonance},
we propose a simple analytic construction
of Wannier-like functions valid at the exact $s-p$ resonance, showing that this leads
to results which capture the essential features of the 
tight binding model obtained within the general method of ref. \cite{Modugno}.
The latter allows us to compute the tunneling amplitudes in a broad range of lattice 
depths also far from $s-p$ resonance, where seemingly the bands are 
decoupled, see Section~\ref{optimal}. Finally, we compare the model obtained with 
different propositions discussed in the literature. 

\section{Model}
\label{model}

For simplicity we consider a purely one dimensional (1D) case with the 
Hamiltonian describing spinless atoms with contact interactions (described by effective coupling constant $g$)
confined in the optical lattice potential $V(x)$:
\begin{equation}
\hat{\cal H} = \int\!\!\mathrm{d}x \hat\Psi^\dagger \left[-\frac{\hbar^2}{2M}\frac{\partial^2}{\partial x^2}
+ V\right] \hat\Psi+
g\int\!\!\mathrm{d}x \hat\Psi^{\dagger 2}\hat\Psi^2. 
\label{h1}
\end{equation}
The field operator $\hat\Psi(x)$ annihilates a~boson at $x$. $\hat\Psi(x)$ is expanded in a
single-particle basis of localized wave functions $f_i^{\alpha}(x)$:
\begin{equation} \label{PSI}
\hat\Psi(x) = \sum_i\sum_{\alpha} \hat{a}^{\alpha}_i f_i^{\alpha}(x)
\end{equation}
The operator $\hat{a}^{\alpha}_i$ annihilates a~boson   at site $i$, while the $\alpha$ index numbers the 
Bloch bands of the periodic potential $V(x)$. Performing the
 integrals in (\ref{h1}) results then in a (multiband) tight binding representation of the problem. 
We shall limit ourselves to relatively weak interactions for which the best localized basis
 states are single particle states determined only by the shape of the potential.
We remark that in the limit of
 strong interactions the single particle basis may not be the optimal one 
\cite{Johnson09,Mering2011,Bissbort2012,Luhmann2012,Lacki2013a}. Actually, despite its
recent progress, the construction of a tight binding model for strong
 interactions remains an open problem.

The period-two SL potential is assumed of the form
\begin{equation}
 V(x)=V_0\left[\sin^2(k_L x+\phi_1)+\epsilon \sin^2(2k_L x+2\phi_2) \right],
 \label{eq:potential}
\end{equation}
with $V_0$ being the amplitude and $k_L$ the wavevector of the laser beam. The case $\phi_1=\phi_2$  
represents a rigid shift of the whole potential.
 Then the sign of $V_0$ as well as the parameter $\epsilon$ (assumed positive) determines the 
 potential shape. $V_0<0$ corresponds to a
 collection of wells with the same minima and alternating heights of the barriers separating 
 them 
\cite{Bloch13,Grusdt13}.
Note that the period of the potential  (\ref{eq:potential}) is $d=\pi/k_L$, and the primitive 
cell contains a double well.
 Putting $\phi_1\ne\phi_2$ allows to modify both the minima and the maxima of the potential 
 simultaneously \cite{Bloch13}. 
Positive values of $V_0$  for $\phi_1=\phi_2$ yield to wells of alternating depths enabling, 
for an appropriately chosen $\epsilon$, an efficient
 coupling between $s$-type states in the shallower wells and $p$-orbitals in the deeper wells. 
 A similar scheme was used in \cite{hemmerich} to effectively
 populate $p$-orbitals in a two-dimensional lattice.

Let us come now to the choice of suitable single-particle localized states $f_i^{\alpha}(x)$ 
in Eq. (\ref{PSI}), which is the subject of this work. 
Generalized Wannier functions are expressed in terms of Bloch eigenfunctions as
\begin{equation}
w_{j\alpha}(x)=\sqrt{\frac{d}{2\pi}}
\int_{\cal{B}} \!\!dk ~e^{-ikja}\sum_{\beta=1}^{N}U_{\alpha\beta}(k)\psi_{\beta,k}(x)
\label{eq:mlwfs}
\end{equation}
where ${\cal{B}}$ is the first Brillouin zone. The unitary
matrices must be continuous and periodic
functions of $k$ in order to preserve the Bloch theorem. For a simple, say sinusoidal  $V(x)$, the optimal basis 
is composed of exponentially localized single-band ($U_{\alpha\beta}(k)=e^{i\theta_{\alpha}(k)}\delta_{\alpha\beta}$)
Wannier functions \cite{kohn}. 
However, when the elementary cell contains two minima, as in the present case,
single particle Wannier states are generally not localized in a single potential minimum. Besides, 
in the present work we shall choose the parameters $\epsilon$ and $V_0$ such that the lowest
band is well separated from the others, with the next 
two lying very close and we will cosider the dynamics of the first and second excited bands.
Consequently, as exposed in \cite{Modugno}, we will consider the mixing of the relative two
Bloch levels through a choice of the mixing matrix $U_{\alpha\beta}(k)$ built to minimize
the spread of the corresponding Wannier functions. This procedure yields a localized $s$-like
state in the shallow well and a $p$-like state in the deeper well.
This configuration with two close lying bands is interesting owing to the fact that one may 
effectively populate the $p$-orbital (like in the two-dimensional experiment
 \cite{Olschlager13}). This situation is also relevant for the realization of an effective
 Dirac dynamics with ultracold atoms in bichromatic optical 
lattices \cite{witthaut,salger,lopez-gonzalez}.

To proceed,  we use first the standard approach for periodic systems, i.e. diagonalization of the 
single-particle Hamiltonian as 
expressed in the recoil units of $E_R=\hbar^2k_L^2/2M$. Thus, in the following, energies 
(and the energy parameters such as the barrier 
height $V_0$) will be given in units of $E_R$ and the corresponding convenient unit of 
length will be $k_L^{-1}$. Also, from now on, we set
 $\phi_1=\phi_2=0$ in (\ref{eq:potential}), corresponding to a potential configuration with degenerate maxima. 
 The periodic part $u_{nk}(x)$ of the Bloch waves, $\psi_{nk}(x)=e^{ikx}u_{nk}(x)$, are obtained from a 
 standard diagonalization procedure in Fourier space. Notice that they can be affected by arbitrary, 
uncorrelated phase factors for different quasimomenta $k$. Then, in order to obtain $u_{nk}(x)$ - 
continuous functions of $k$, we need to fix all the phases
 to zero at some point in $x$ space \cite{Modugno} or equivalently, for a real matrix diagonalization 
 (as possible for our 
simple potential) to choose the same signs of $u_{nk}(x)$ say at $x=0$.

\section{$s-p$ resonance}
\label{resonance}

\begin{figure}[b]
\includegraphics[width=0.8\columnwidth]{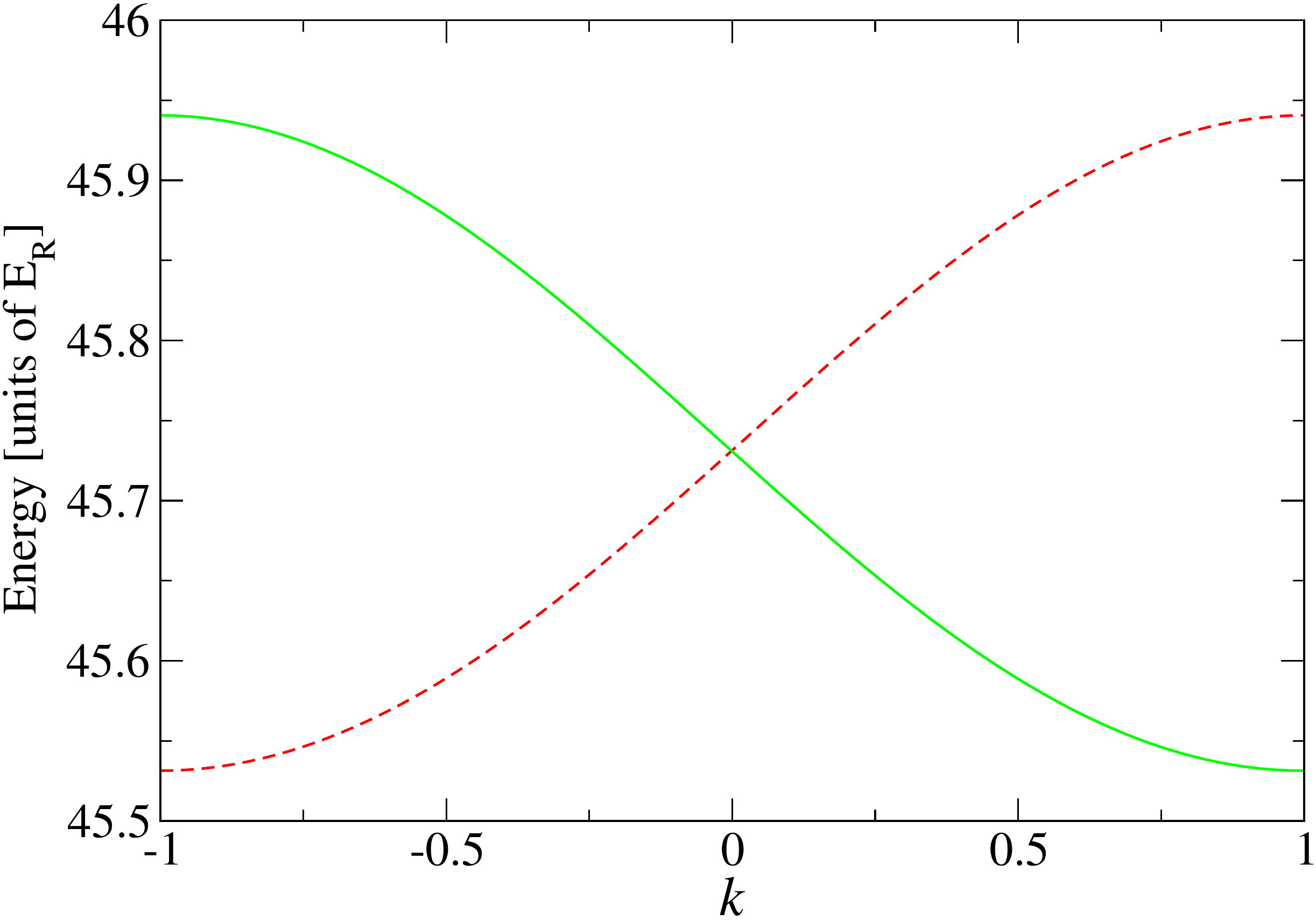}
\caption{Ccolor online) Energy bands for the double-well potential (\ref{eq:potential}) with parameters $V_0=32E_r$, 
$\epsilon=2$ and $\phi_1=\phi_2=0$. The lowest energy band is not shown, the red (dashed) second band crosses with
 the green (solid) third band leading to a linear ``relativistic'' dispersion around $k=0$}
\label{fig:bands}
\end{figure}

\begin{figure}
\includegraphics[width=\columnwidth]{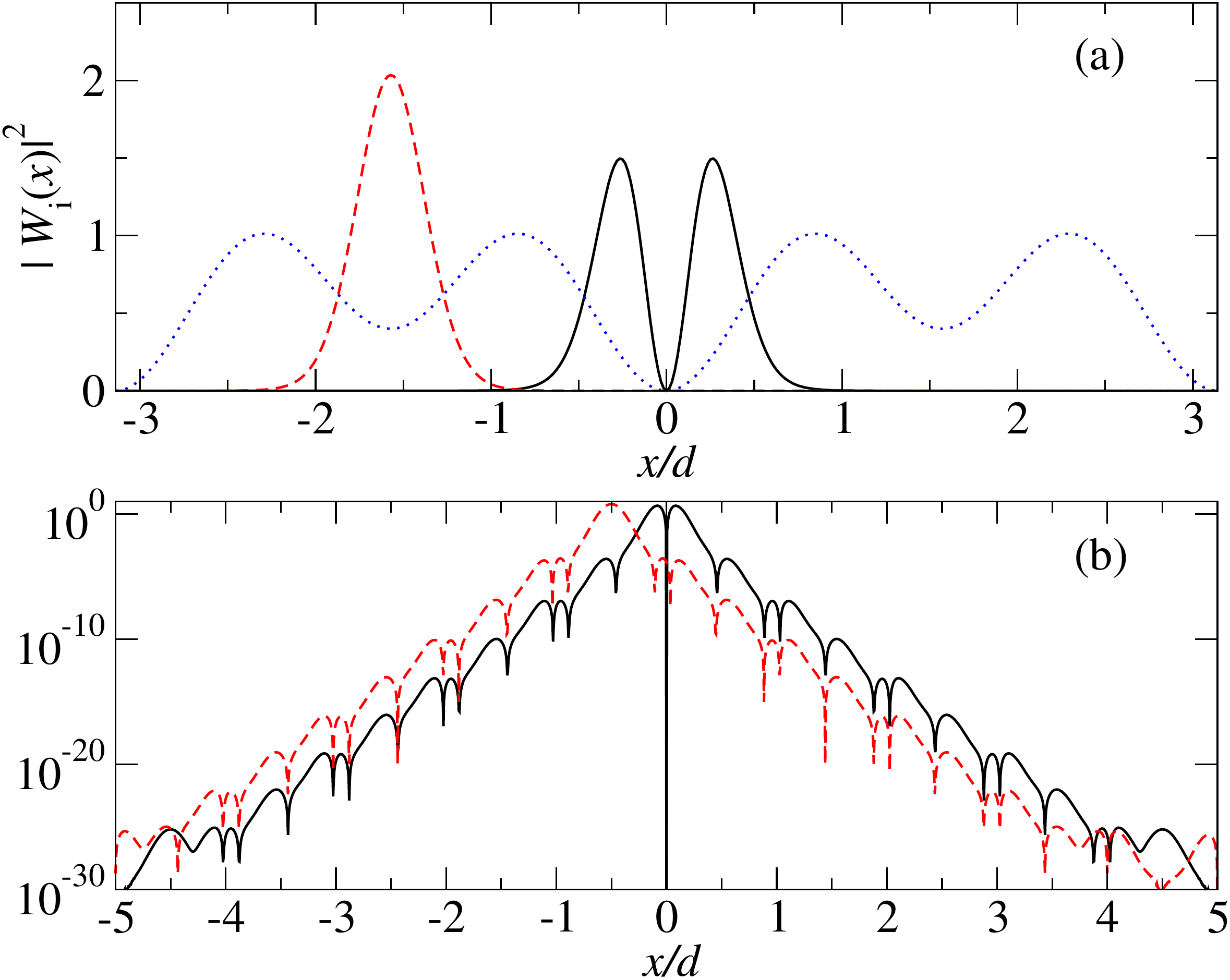}
\caption{(Color online) Top panel: two localized squared Wannier 
functions of the double-well potential (\ref{eq:potential}) with parameters 
$V_0=32E_r$, $\epsilon=2$ and $\phi_1=\phi_2=0$. Functions $|W_s|^2$ (red dashed line) 
and $|W_p|^2$ (black solid line) come from Bloch functions transformed by (\ref{eq:u}). 
Dotted blue line depicts the lattice potential (not in scale). Bottom (b) panel shows the same 
Wannier functions in the logarithmic scale revealing their exponential localization within
double precision arithmetics.}
\label{fig:loc}
\end{figure}

Let us examine the resonant case. To this end, we take a particular $V_0$ and by diagonalizing 
the one-particle Hamiltonian for different
 values of $\epsilon$ we choose the one which makes the $2$nd and $3$rd Bloch bands cross (see Fig.~\ref{fig:bands}).  
Then, by making a harmonic approximation for the wells, it is easy to find that the condition of 
resonance is  $\epsilon=V_0/(16E_R)$.
 Quite surprisingly, we have found numerically that this condition holds to 
 a very good approximation also for relatively shallow wells, up to $V_0=16E_R$. 
Again, we remark that the use of the single-band approach (namely 
$U_{\alpha\beta}(k)=e^{i\theta_{\alpha}(k)}\delta_{\alpha\beta}$ in Eq. (\ref{eq:mlwfs})) 
gives rise to Wannier functions $W_{0i}$ ($i=1,2,3$) well localized in a single well only for 
the states corresponding to isolated bands \cite{Modugno}. For the present configuration, this 
is the case of $W_{01}$, whereas $W_{02}$ and $W_{03}$ have non-zero contributions both in the 
deeper and shallower wells (at resonance). 
Consequently, a non trivial mixing of the 2nd and 3rd bands is necessary.  Remarkably, in this particular case,
optimally localized states can be built from a simple analytic formula.
In fact, let us consider the following Wannier-like states
\begin{equation}
 W_n(x-R)=\frac{1}{\sqrt{2}}\int_{\mathcal{B}}dk e^{-ikR}\psi_{nk}(x),
\label{single}
\end{equation}
with $\mathcal{B}$ standing for the first Brillouin zone, so in our case: $d=\pi$ and $k\in [-1,1]$. 
This formula leads to the standard definition of the Wannier 
functions only when $R$ is a vector of the Bravais lattice, namely $R=R_{j}=j\pi$.
Then, by using the following transformation 
\begin{equation}
 U=\frac{1}{\sqrt{2}}\left(\begin{array}{cc}
1 & 1 \\ 
- i &  i
\end{array}\right)
\label{eq:u}
\end{equation}
acting on the $n=2,3$ Bloch functions, and taking $R=(2j-1)\pi/2\equiv R^{(s)}_{j}$ for $s$-states and $R=j\pi\equiv R^{(p)}_j$ for $p$-states, 
we write (for the cell at $j=0$) 
\begin{eqnarray}
 \label{eqn:wans}
W_s(x)&=&\frac{1}{2}\int^{1}_{-1}\!\!\!dke^{+ik\frac{\pi}{2}}\Big[\psi_{2k}(x)+\psi_{3k}(x) \Big],  
\\  \label{eqn:wanp}
W_p(x)&=&\frac{i}{2}\int^{1}_{-1}\!\!\!dk\Big[-\psi_{2k}(x)+\psi_{3k}(x) \Big]. 
\end{eqnarray}
Note that this transformation, along with the definition of the initial gauge for the Bloch states 
(see end of Sect. \ref{model}), univocally defines this sets of Wannier functions.
The functions $W_{s}$ and $W_{p}$ are localized in the minima of the shallow well ($x=-\pi/2$) and the 
deep one ($x=0$) respectively, see Fig. \ref{fig:loc}(a)
\cite{foot1} . Both exhibit clear exponential 
fall-off, as seen in Fig. \ref{fig:loc}(b). For other cells 
the translation by $x=\pi$ (with integer $j$) applies, as required by the Bloch theorem.
As it will be discussed in Sect. \ref{optimal}, this simple transformation captures the essential features of the general method of Ref. \cite{Modugno} in describing the model and it leads to simple analytic expressions for the tunneling coefficients.

Then we introduce the following shorthand notation $W_s(x-R_{j})=\langle x | s_j\rangle$ and   
$W_p(x-R_{j})=\langle x |  p_j\rangle$ for the 
Wannier-like functions belonging to the $j$-th cell, and denote with $E_n(k)$ the energy bands as 
indicated in Fig. \ref{fig:bands}. Then, it 
is straightforward to find the following expressions for the tunneling coefficients

\begin{eqnarray}
J_{r}&=&-\langle s_j | \hat{H}_0 | p_{j} \rangle=\frac{1}{2\pi}\int_0^{\pi} dk (E_3(k)-E_2(k))\sin\left({k}/{2}\right), 
\nonumber \\
J_{l}&=&-\langle p_{j-1} | \hat{H}_0 | s_{j}\rangle= - {J_r},\nonumber \\
J_s&=&-\langle s_j | \hat{H}_0 | s_{j+1} \rangle=-\frac{1}{2\pi}\int_0^{\pi} dk (E_2(k)+E_3(k))\cos(k), \nonumber\\ 
J_p&=&-\langle p_j | \hat{H}_0 | p_{j+1} \rangle=J_s. 
\label{eq:tuncoef1}
\end{eqnarray}

With the choice of functions (\ref{eqn:wans}-\ref{eqn:wanp}) the tunneling amplitudes are real. 
Notice that the tunneling amplitude, $J_r$ from $s$ $i$-site to the right (to $p_{i+1}$) 
has opposite sign to that corresponding to left hand side hop $J_l$ (due to the asymmetric 
character of the $p$ orbital). The tunneling amplitudes  $J_{s,p}$ -- between the same flavors 
$s$ and $p$ coincide at  $s-p$ resonance.

One may adjust arbitrarily phases of Wannier functions. Changing the sign of every second 
double site (i.e. both $s$ and $p$ orbital functions) one may realize the ``standard'' system 
with $t$ amplitudes for nearest neighbor hopping and $t'$ for next-nearest hopping (called often
$t-t'$ or $J_{1}-J_{2}$ model). This procedure, however, breaks partially the translational 
invariance of the system. To recover the translational invariance, the elementary cell has 
to be extended to four sites. Such a change of phases affects also the dispersion relation.

\section{A general approach}
\label{optimal}

As we have anticipated, the analytic transformation discussed in the previous section is designed specifically for the resonant case and cannot be applied to a generic situation. In the latter case, a general approach  
is represented by the so-called maximally localized Wannier functions (MLWFs) introduced by Marzari and Vanderbilt \cite{Marzari97, Marzari12}. They are obtained by generalizing the definition of Wannier functions for the case of a set of (almost) degenerate Bloch bands, and then minimizing their spread by means of an appropriate gauge transformation of the Bloch functions.
By construction, this transformation is differentiable and is defined as periodic, in order to keep the periodicity of the Bloch functions. This method has been adopted by two 
of us for the lowest bands of 1D superlattice potential  \cite{Modugno}. It can also  be 
applied straightforwardly to the present situation, yielding a set of MLWFs both at the $s-p$ resonance 
and in its vicinity, or in any situation in which there are two almost degenerate bands (well
separated from the others). While the details of the approach have been presented in 
\cite{Modugno}, here we briefly recall that it consists in using a specific two-step gauge 
transformation that makes vanishing the diagonal 
and off-diagonal gauge-dependent terms of the Wannier spread. Such a transformation is obtained 
by numerically solving a set of ordinary differential equations, with suitable boundary conditions.
As a specific example, in the following we will fix $V_{0}=32E_{R}$ (for which the $s-p$ resonance occurs at $\epsilon=2$) to illustrate the typical behavior in the tight-binding regime.

Let us first discuss the resonant case, by comparing the general method with the results
obtained by means of the analytic ansatz. First, observe that since the bands are degenerate at $k=0$ 
(see Fig.~\ref{fig:bands}), 
the states in Eqs. (\ref{eqn:wans}-\ref{eqn:wanp}) are built assuming that the second and
the third bands are swapped for $k>0$,  thus loosing the periodicity of the conventional Bloch bands. 
Alternatively, one could consider symmetric bands touching at $k=0$, with an additional $sgn(k)$ factor 
in the lower row of the mixing matrix, 
abandoning the periodicity and the continuity of the unitary transformation.
 This is a consequence of the fact that the specific choice of term $R$ in Eq. (\ref{single}) 
 for $s$ and $p$-like states is, as a matter of fact,  equivalent 
to considering an effective lattice with half periodicity, that includes a single minimum in the 
elementary cell (notice that we are considering 
the case of degenerate maxima by having set $\phi_1=\phi_2=0$ in
Eq. (\ref{eq:potential})).  
Remarkably, the Wannier-like states in Fig. \ref{fig:loc} are almost coincident with those of the MLWFs obtained from the
general procedure of Ref. \cite{Modugno}. Also the corresponding tunneling amplitudes are well captured, as we will show in the following.  

\begin{figure}
\centerline{
\includegraphics[width=\columnwidth]{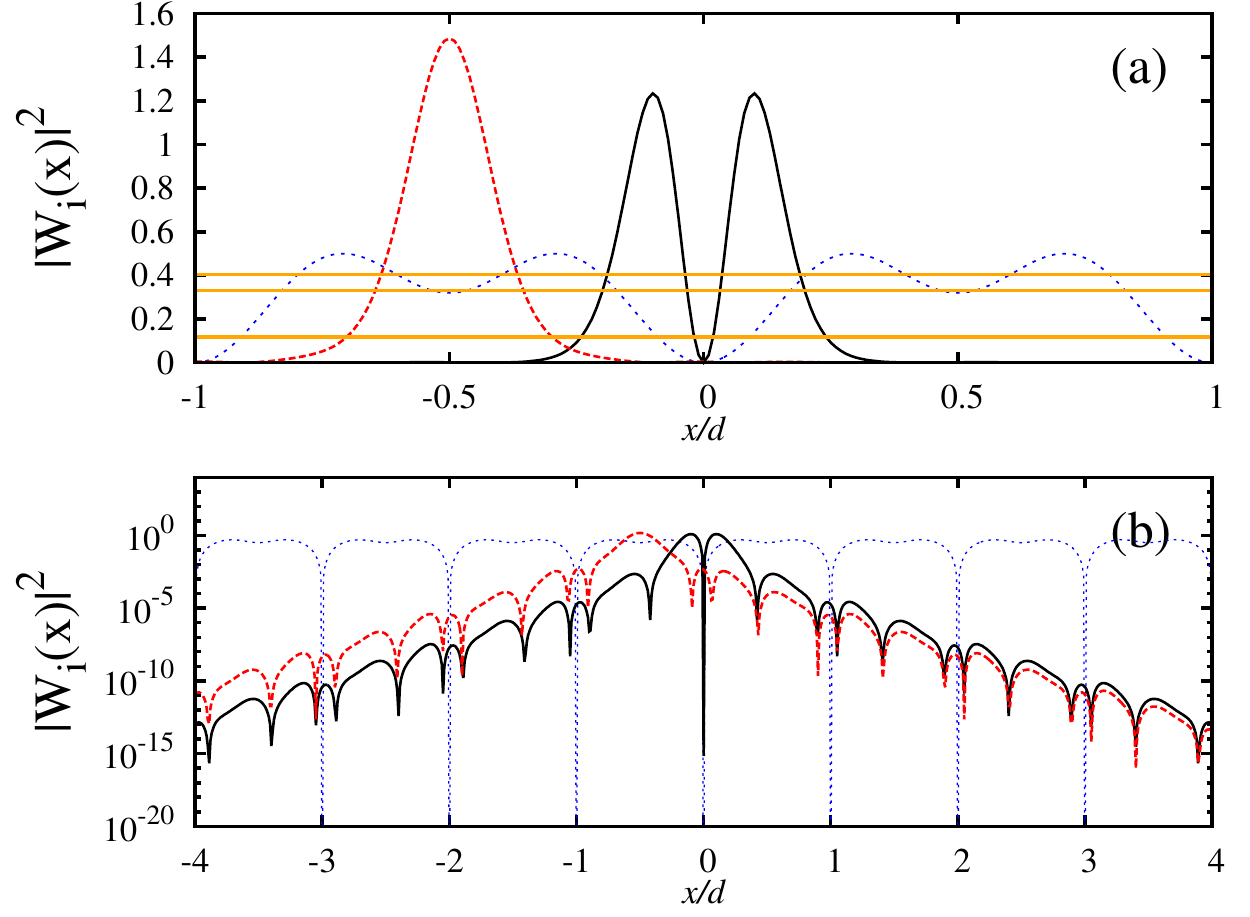}
}
\caption{(Color online) (a) Density plot of the $s-$ (red dashed line) and $p-$like
 (black solid line) MLWFs for $\epsilon=1$, corresponding to a large detuning from the resonance.
 The potential is represented by the dotted (blue) line, whereas the horizontal orange stripes represents the Bloch bands
 (on the same scale as the potential). (b) The same composite-band MLWFs are shown here in
 logarithmic scale. Note the exponential decay of the tails.}
\label{fig:wan-off1}
\vspace{0.5cm}
\centerline{
\includegraphics[width=\columnwidth]{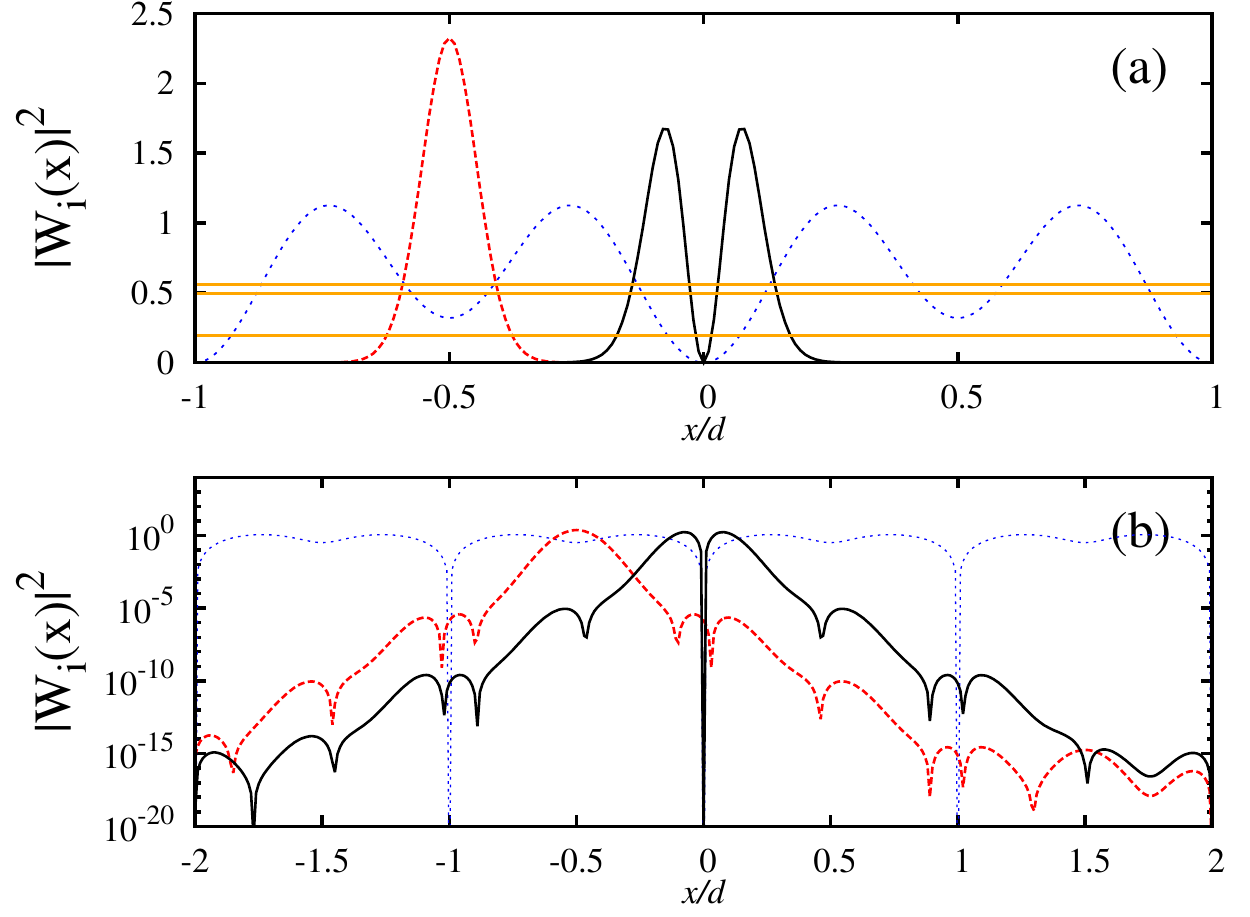}
}
\caption{(Color online) Same as above, but for $\epsilon=3$.}
\label{fig:wan-off3}
\end{figure}

\begin{figure}
\includegraphics[width=0.95\columnwidth]{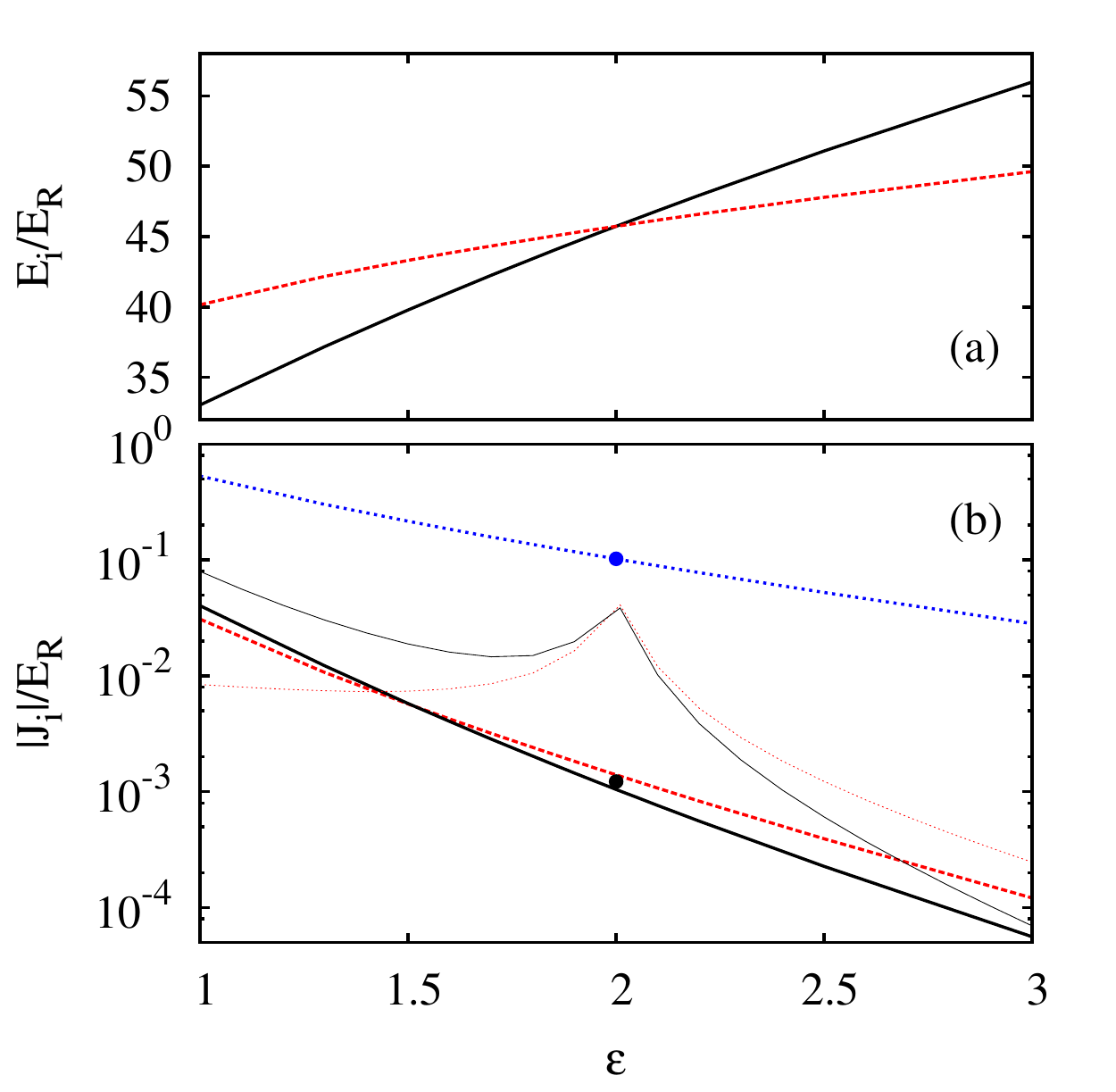}
\caption{(Color online) 
Onsite energies and tunneling amplitudes obtained from the composite-band MLWFs approach as a function of $\varepsilon$.
(a) Onsite energies $E_{s}$ (black solid line) and $E_{p}$ (red dashed line).
(b) Tunneling amplitudes: the blue (dotted) line corresponds to the amplitude $|J_{l}|=|J_{r}|$ for nearest-neighbor tunneling,  i.e. between $s$ and $p$ orbitals; the black (solid) line corresponds to the $s-s$ tunneling amplitude $J_{s}$, while the red (dashed) line represents the $p-p$ tunneling $J_{p}$. Thin lines represents the values of $J_{s}$ and $J_{p}$ obtained from a single-band calculation (see text). The (blue and black) dots correspond to the prediction of the analytic expressions  in Eq. (9) for the resonant case ($\varepsilon=2$).}
\label{fig:tun-32}
\end{figure}

Let us now turn to the off-resonant case. As an example, in Figs. \ref{fig:wan-off1} and \ref{fig:wan-off3} we show the  MLWFs for $\epsilon=1,3$, respectively. Each wave function is shown both in linear and logarithmic scale, in order to make evident that each of them is well localized around a single site, and to show the exponential falloff of their tails. Then, in Fig. \ref{fig:tun-32}a we show the behavior of the two onsite energies, namely  $E_{s}\equiv\langle s_{j}|\hat{H}_{0}|s_{j}\rangle$ and $E_{p}\equiv\langle p_{j}|\hat{H}_{0}|p_{j}\rangle$, as a function of $\epsilon$. Note that below resonance $E_{s}>E_{p}$, whereas the situation is reversed above the resonance. The two on-site energies become degenerate exactly at $\epsilon=2$.

Similarly, in Fig. \ref{fig:tun-32}b we show the behavior of the various tunneling amplitudes. They are characterized by a monotonic decrease with increasing $\epsilon$, as a consequence of the deepening of the potential wells. Note also that in this case the degeneracy between the $s-s$ and $p-p$ tunneling amplitudes takes place around $\epsilon=1.5$ and not at the resonance, as the tunneling rates are determined by the specific form of the Wannier functions, and not just by the value of the onsite energies. This is different from the result of the analytic approach, that predicts an exact degeneracy between $J_{s}$ and $J_{p}$ at the resonance (see the last equation in (\ref{eq:tuncoef1})).
Nevertheless, apart from this detail, the analytic approach essentially captures 
the correct values of the tunneling coefficient at resonance. We also remark that, due to parity of $p$ orbital involved in $s-p$ hopping, the nearest neighbor tunneling amplitudes 
from a given site to the left and to the right have the same magnitude but  opposite sign, namely $J_{l}=-J_{r}$, as already obtained for the analytic case discussed in the previous section. 

Remarkably, that in the whole range of $\epsilon$ considered here, the amplitude of the nearest neighbor tunneling $J_{l/r}$ is an order of magnitude larger than the next-to-nearest neighbors tunneling amplitudes $J_{s}$ and $J_{p}$. Naively, this behavior is expected as the former corresponds to the hopping between neighboring wells. However, far from the resonance (e.g. at $\epsilon\simeq1$ or $\epsilon\simeq3$) one may also expect the usual single-band approach to provide Wannier functions that are well localized around each potential minima \cite{Modugno}. Indeed, this is the case, as it will be shown in the following section. In this framework, the $s-$ and $p-$like Wannier states are orthogonal (they belong to different Bloch bands), and the nearest neighbor tunneling is exactly vanishing. Then, in order to clarify this apparent paradox, in the next section we will work out a thorough comparison between the single- and composite-band approaches.

\section{Far from the resonance: comparison with the single-band approach}

Let us recall that proper single-band maximally localized Wannier functions can be constructed with the same approach \textit{a la} Marzari and Vanderbilt discussed in the previous section. In this case it is necessary to minimize just the diagonal spread, as the off-diagonal one is vanishing by construction. In practice, this can be achieved by solving a set of simple ODEs for the phase $\phi_{n}(k)$ of the Bloch functions, corresponding to a $U(n)$ transformations ($n$ being the number of Bloch bands one is interested in) \cite{Modugno}. 
Remarkably, the values of the phases $\phi_{n}(k)$ obtained from the numerical solution allow for a simple analytic approximation that consists in using Eq. \eqref{single} for $n=2,3$ and adjusting accordingly the centers $R$ of the Wannier functions to the bottoms of the corresponding wells, with an additional factor $sgn(k)$ for the $p$-state (that is centered in $x=0$). Indeed such a procedure, far from the resonance, leads to exponentially localized Wannier functions that are indistinguishable from those obtained numerically by minimizing their spread. We remark that this construction requires that the initial gauge for the Bloch functions is that defined at the end of Sec. \ref{model}, and that each band is associated to the appropriate well. In particular, the $p$-type Wannier function corresponds to $n=2$ for $\epsilon$ below the resonance (i.e. $\epsilon<V_0/16$) and $n=3$ above it \cite{foot2}.
\begin{figure}[b]
\includegraphics[width=0.95\columnwidth]{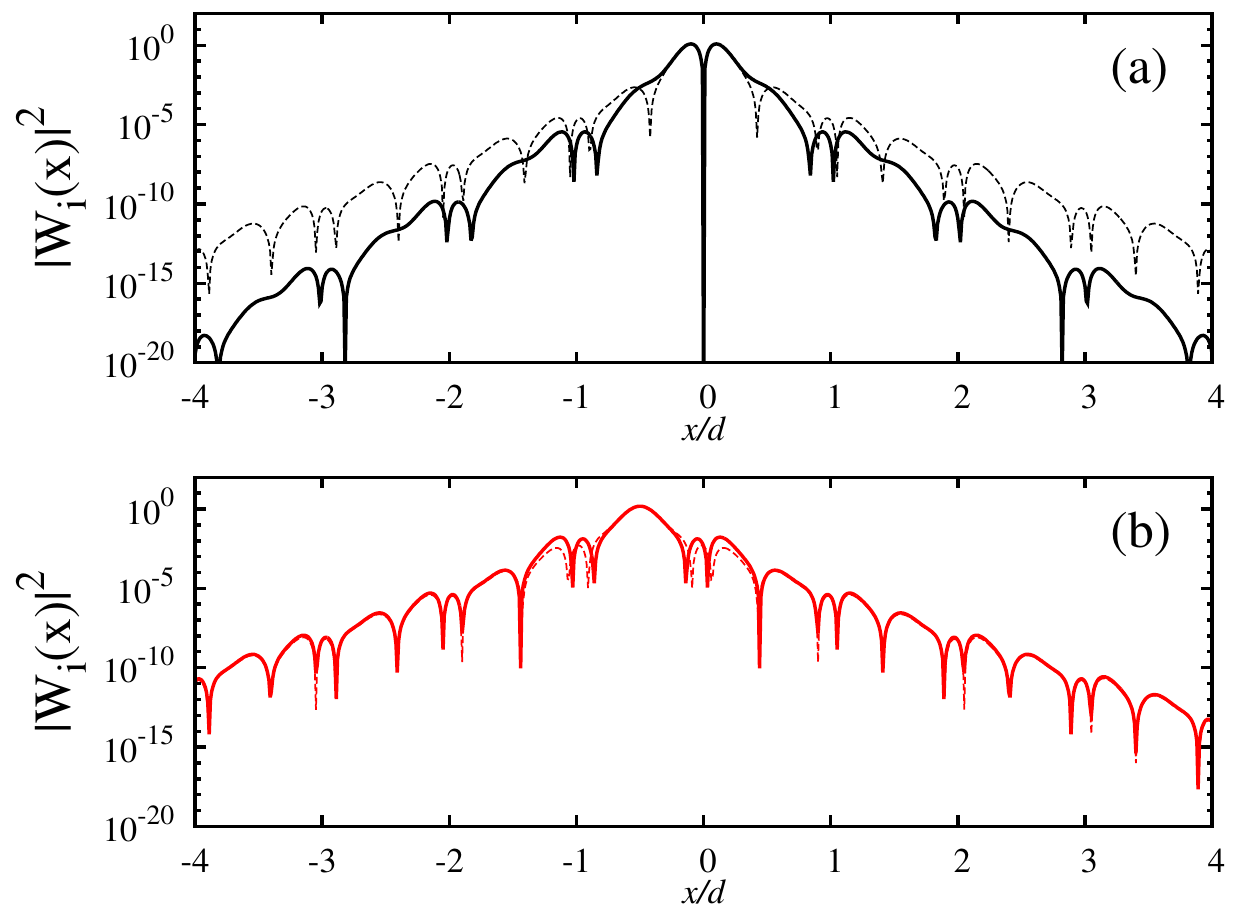}
\caption{(Color online) (a) Density plot of the $p-$like single-band Wannier functions (solid line) for $\epsilon=1$, 
compared with that obtained from the composite-band approach (dashed line). (b) The case of the $s-$like MLWF, as above.}
\label{fig:comparison1}
\end{figure}
\begin{figure}[t]
\includegraphics[width=0.95\columnwidth]{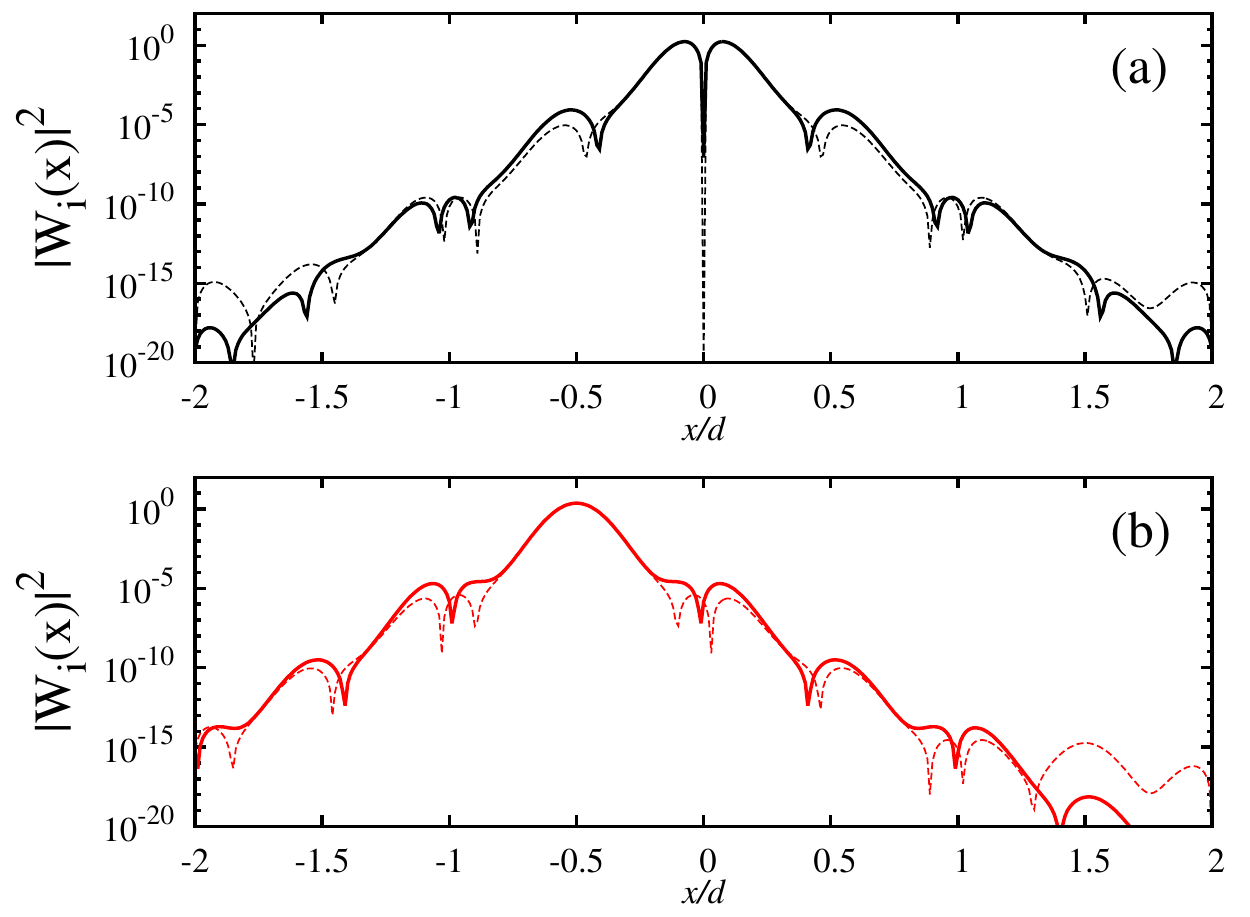}
\caption{(Color online) Same as Fig. \ref{fig:comparison1}, but for $\epsilon=3$.}
\label{fig:comparison3}
\end{figure}

As an example, the single-band Wannier functions obtained for $\epsilon=1,3$ are shown in Figs. \ref{fig:comparison1}, \ref{fig:comparison3} in comparison with those obtained from the composite-band approach.
Specifically, in these figures we show the modulus in logarithmic scale.
In all cases, both sets of MLWFs show the same bulk behavior and are characterized by a nice exponential falloff of the tails. For $\epsilon=1$, where the two Bloch bands are well separated (see Fig. \ref{fig:wan-off3}), the $p-$type Wannier function obtained from the single-band approach turns out to be more localized than that corresponding to composite bands (note the different slope of the exponential envelope), whereas the situation is reversed por the $s-$like function (note the first two side lobes). Instead, for $\epsilon=3$ - here the two bands are closer to each other - the composite-band MLWFs are more localized overall. In any case, we remark that the main difference between the two approaches resides in the structure of the Wannier functions as complex numbers (so far we have been showing just their modulus), that makes e.g. the $s$ and $p$ single-band states orthogonal. This is responsible for the very different structure of the tunneling coefficients, as shown in Fig. \ref{fig:tun-32}b.
As a matter of fact, the single-band approach (that, we remind, is not suitable around the resonance as the Wannier functions occupy both wells \cite{Modugno}) corresponds to a picture with two sublattices one of type $s$  the other of type $p$ (with different tunneling rates within each sublattice), that are decoupled in the absence of interactions.

In the following, we will discuss which are the implications of the two pictures (single or composite bands),
by considering first the structure of the single-particle spectrum and then the role played by interactions.

\subsection{Single-particle spectrum}

Let us recall that the single particle term of the tight-binding Hamiltonian up to nest-to-nearest neighbors takes the form
\begin{align}
\hat{H}_{0} &\simeq 
\sum_{\alpha=s,p}\sum_{j}E_{\alpha}\hat{n}_{j\alpha}
-\sum_{\alpha=s,p}\sum_{j}J_\alpha(\hat{a}_{j\alpha}^\dagger\hat{a}_{(j+1){\alpha}} +h.c.)
\nonumber\\
&
-J\sum_{j}\left(\hat{a}_{js}^\dagger\hat{a}_{jp}
-\hat{a}_{js}^\dagger\hat{a}_{(j-1){p}} +h.c.\right).
\label{eq:fullH}
\end{align} 
where $\hat{a}^{\dagger}_{j\alpha}$ ($\hat{a}_{j\alpha}$) represent creation (annihilation) operators associated to each lattice site, with $\alpha=s,p$.

The single-particle spectrum can be obtained by considering the following mapping to momentum space \cite{Modugno},
$\hat{b}_{k\alpha}=\sum_j e^{i\pi kj}\hat{a}_{j\alpha}/\sqrt{2}$. Then, 
the single-particle Hamiltonian can be written as
\begin{equation}
 \hat{H}_{0}=\sum_{\alpha\beta}\int_\mathcal{B} dk h_{\alpha \beta}(k)\hat{b}^\dag_{k\alpha}\hat{b}_{k\beta},
\end{equation}
with $h_{\alpha \beta}(k)$ defined as
\begin{equation}
h_{\alpha \beta} =\left(\begin{array}{cc}
E_s-2J_s\cos(\pi k) & -J(1-e^{i\pi k}) \\ 
-J(1-e^{-i\pi k}) & E_p-2J_p\cos(\pi k)
      \end{array}\right). 
\end{equation}

For the single-band approach $J\equiv0$, so that the dispersion relation takes the familiar form
\begin{equation}
E_{\alpha}^{sb}(k)=E_\alpha -2J_\alpha\cos(k\pi).
\end{equation}
In the general case, the diagonalization of $h_{\alpha \beta}$ yields the following dispersion relation
\begin{eqnarray}
E_{\pm}^{cb}(k)&=&E_{+}-2J_{+}\cos(k\pi)
\\ \nonumber
&\pm&\sqrt{ \left(E_{-}+2J_{-}\cos(k\pi)\right)^2
+(2J)^{2}\sin^{2}(k\pi/2) },
\end{eqnarray}
with $E_{\pm}\equiv(E_s\pm E_p)/2$, $J_{\pm}\equiv(J_s+J_p)/2$.

\begin{figure}[b]
\includegraphics[width=0.95\columnwidth]{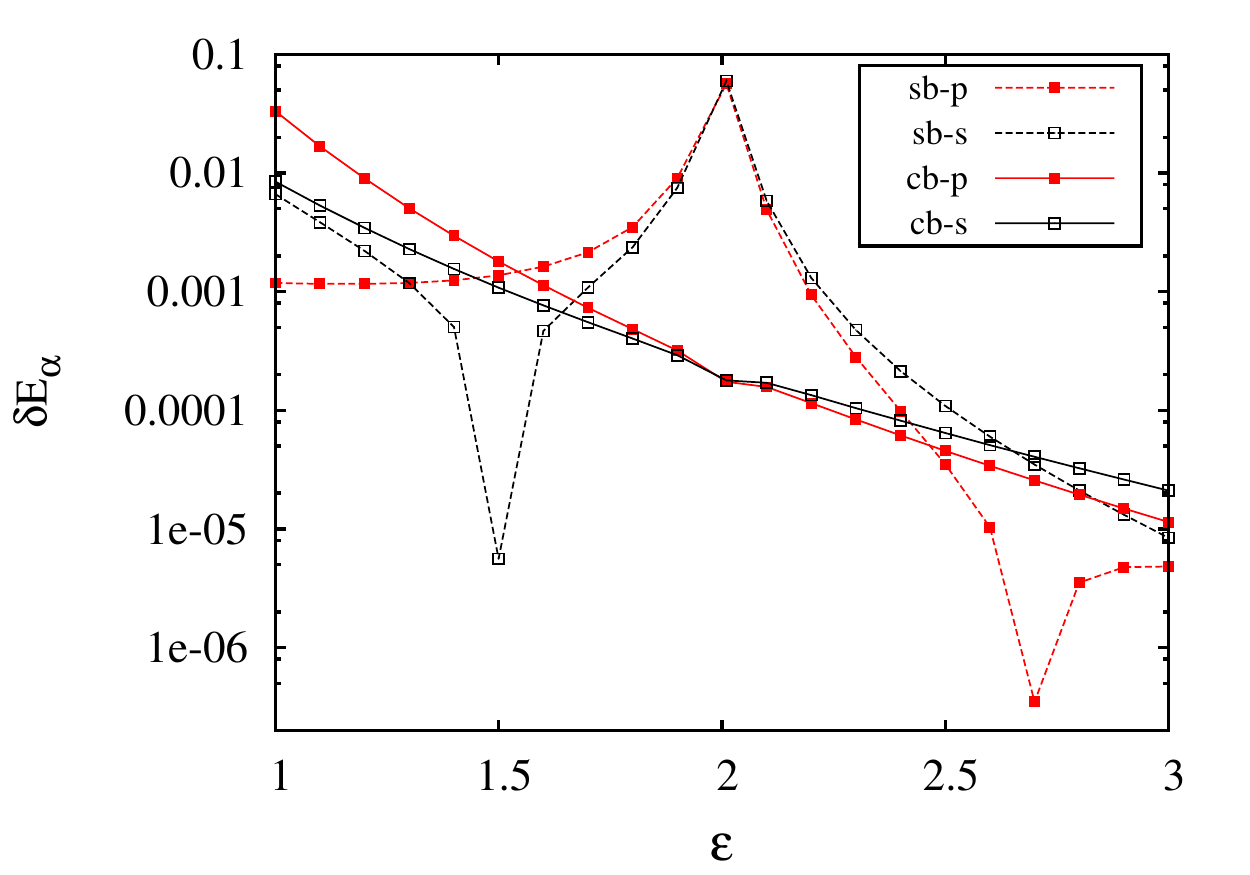}
\caption{(Color online) Plot of $\delta E_{\alpha}$ ($\alpha=s,p$, see text) as a function of $\varepsilon$, as obtained 
from the single- and composite-band approaches (indicated in the label as `sb' and `cb', respectively).
 }
\label{fig:enerdiff}
\end{figure}

Then, the accuracy in reproducing the exact single particle Bloch spectrum  can 
be measured by defining the following quantity \cite{Modugno}
\begin{equation}
\delta E_{\alpha} \equiv \frac{1}{\Delta{E}_{\alpha}^{ex}}\sqrt{\frac{d}{2\pi}\int_{\cal{B}} dk \left[E_{\alpha}(k)-E_{\alpha}^{sb/cb}(k)\right]^{2}}
\end{equation}
that represents the ratio of the quadratic spread between the exact Bloch spectrum $E_{\alpha}^{ex}(k)$ and that obtained from the single- or composite-band approach, to the Bloch bandwidth $\Delta{E}_{\alpha}^{ex}$. 
The behavior of $\delta E_{\alpha}$ as a function of $\varepsilon$ is shown in Fig. \ref{fig:enerdiff}.
This figure deserves a careful analysis. Close to $\varepsilon=1$, the nearest-neighbor approximation works accurately only in the single-band approach, and just for the lowest band (of type $p$). In the other cases,
the inclusion of next-to-nearest tunneling terms would be required \cite{Modugno}. 
Actually, in that limit the difference between the exact Bloch bands and those calculated with the nearest-neighbors tight-binding approach would be visible by eye.
Then, it is evident starting from $\varepsilon=1.5$, the composite-band approach provides an increasing, very good accuracy in reproducing the single particle spectrum, thanks to the fact the the system enters a full tight-binding regime. Instead, the single-band approach is characterized by a mixed behavior, that is dramatically affected by the proximity of the resonance point, where it definitely fails. Only for $\varepsilon\gtrsim2.2$ the single-band approach recovers a good accuracy level.

\subsection{Interaction terms}

\begin{figure}
\includegraphics[width=\columnwidth]{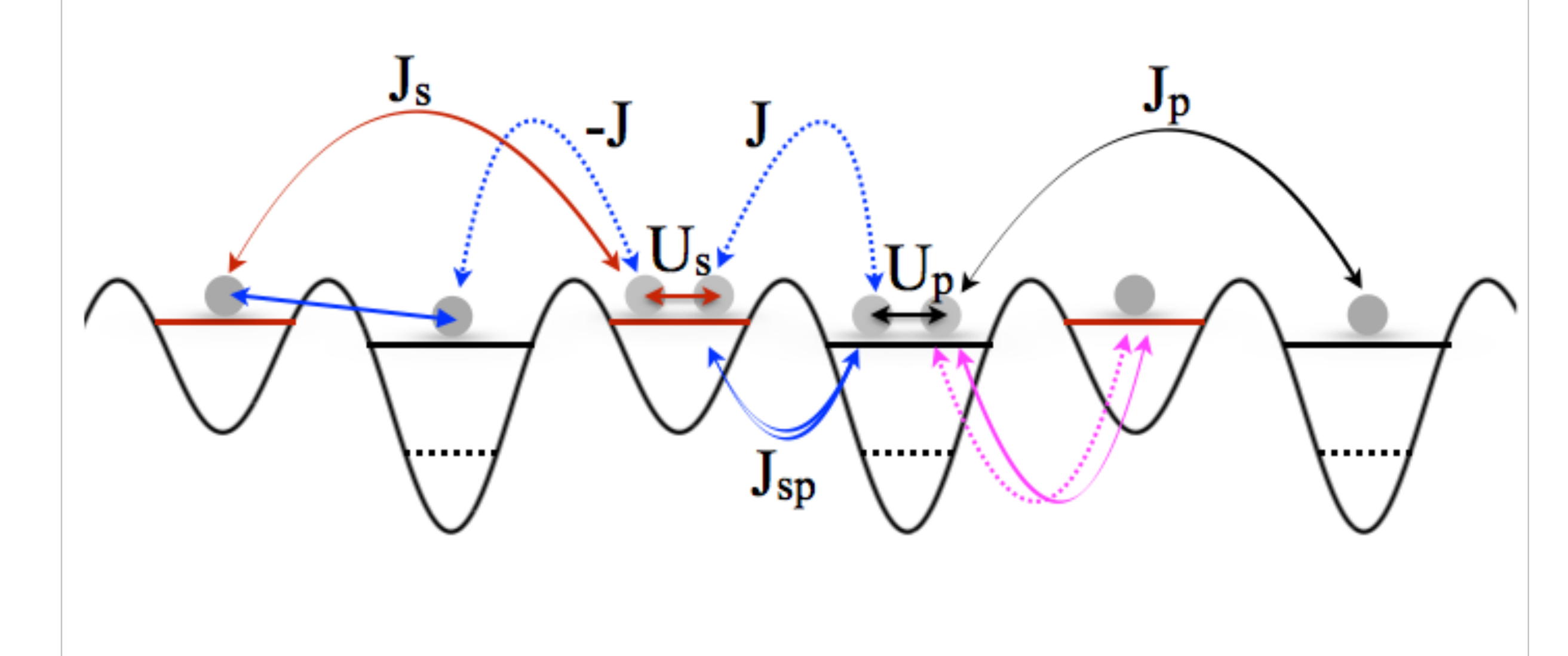}
\caption{(Color online) A sketch of the various tunneling and interaction terms of the extended Bose-Hubbard model.
The color code is the same as that in Figs. \ref{fig:tun-32} and \ref{fig:integrals}. $J_{sp}$ schematically denotes two types of density dependent tunnelings, mediated by $s$ and $p$ orbitals densities. Remember that the nearest-neighbor tunneling coefficient $J$ is strictly vanishing within the single band approach.}
\label{fig:sketch}
\end{figure}

Let us now turn to the interacting part of the many-body Hamiltonian. The full expression reads
\begin{align}
\hat{H}_{int} &= \frac{g}{2}\sum_{\{\alpha_{i}\}=s,p}
\sum_{\{j_{i}\}}\hat{a}_{j_{1}\alpha_{1}}^\dagger\hat{a}_{j_{2}\alpha_{2}}^\dagger\hat{a}_{j_{3}\alpha_{3}}\hat{a}_{j_{4}\alpha_{4}}\cdot
\\
&\cdot\int dx W_{j_{1}\alpha_{1}}^\ast (x) W_{j_{2}\alpha_{2}}^\ast(x) 
W_{j_{3}\alpha_{3}}(x) W_{j_{4}\alpha_{4}}(x),
\label{eq:Hint-full}
\end{align}
with $g$ being the coupling constant. The leading term is represented by usual Bose-Hubbard on-site interaction, namely
\begin{equation}
\hat{H}_{onsite}= \frac{1}{2}\sum_{\alpha=s,p}U_{\alpha}\sum_{j\alpha}\hat{n}_{j\alpha}\left(\hat{n}_{j\alpha}-1\right)
\label{eq:Hint-onsite}
\end{equation}
with $U_{\alpha}=g\int\!d x \left|W_{j\alpha}(x)\right|^4$. In addition to this, here we will consider also
 the effect of next-to-leading terms that couple nearest neighboring wells, i.e. $s$ and $p$ wells. They include a density-density interaction term
\begin{equation}
\hat{H}_{dens-dens}=\frac{g}{2} I_{2s2p}\cdot\hat{n}_{js}\hat{n}_{jp},
\end{equation}
a density induced tunneling
\begin{equation}
\hat{H}_{dens-tun}= \frac{g}{2} I_{1s3p}\hat{a}_{js}^\dagger\hat{n}_{jp}\hat{a}_{jp} + 
\frac{g}{2} I_{3s1p}\hat{a}_{js}^\dagger\hat{n}_{js}\hat{a}_{jp} + (s\leftrightarrow p),
\end{equation}
and the tunneling of pairs
\begin{equation}
\hat{H}_{pair-tun}= \frac{g}{2} I_{2s2p}\cdot\hat{a}_{js}^\dagger\hat{a}_{js}^\dagger\hat{a}_{jp}\hat{a}_{jp} + h.c.
\end{equation}
In the previous expressions we introduced an intuitive notation for the integration integral within the $j$-cell indicating how many Wannier function of a given type enter into the integral
 \begin{equation}
I_{iskp}\equiv\int dx (W_{js}(x))^{i} (W_{jp}(x))^{k},
\end{equation}
taking into account that the Wannier functions can be chosen as real \cite{Brouder}. Notice that there are also similar terms that couple $s-$states in the $j-th$ cell with $p-$states on the left, that is the $(j-1)$th cell. Observe, however, that the sign of the density dependent tunneling ``to the left'' is opposite to that ``to the right'' due to the asymmetry of the $p$-orbital, in parallel with the standard tunneling.
A sketch of the various tunneling and interaction terms of the extended Bose-Hubbard model is shown in Fig. \ref{fig:sketch}.

\begin{figure}
\includegraphics[width=0.95\columnwidth]{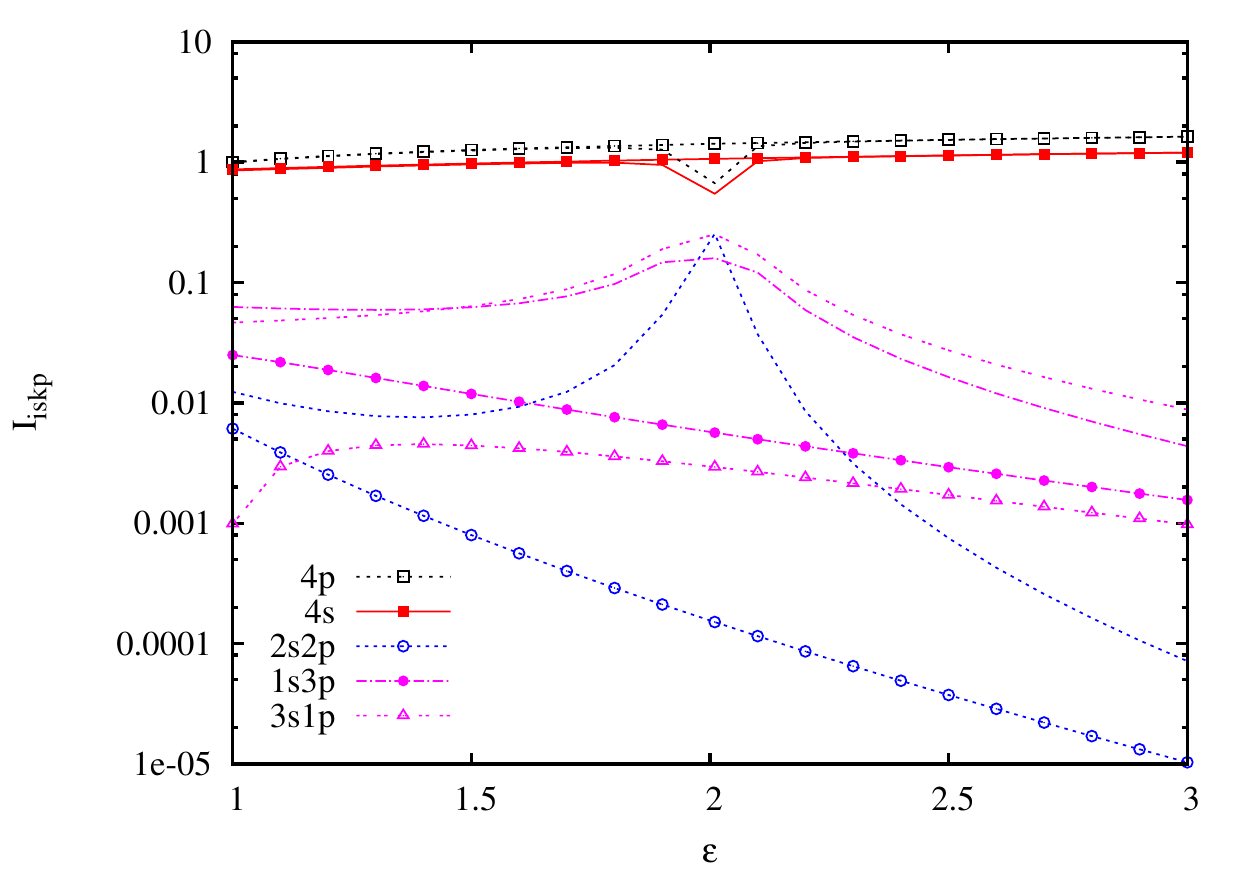}
\caption{(Color online) Plot of the (modulus of) the integrals $I_{iskp}$ characterizing the amplitude of various interaction terms of the extended Bose-Hubbard Hamiltonian. Lines with symbols correspond to the two-band approach, whereas plain lines refer to the single band case. Observe that the two-bands approach yields generally much smaller corrections to standard Bose-Hubbard terms, showing superiority over a single-band Wannier functions.}
\label{fig:integrals}
\end{figure}
The relative weight of the various interaction terms listed above is shown in Fig.~\ref{fig:integrals} where we present the behavior of the various integrals (in modulus) entering in their definition, as obtained from the both the single- and composite-band approaches
 (note that $I_{2s2p}=I_{2p2s}$). This figure shows that, as far as interactions are concerned, the composite-band model outperforms the single-band one, as the next-to-leading terms are significantly smaller in the former case. Notice also that (out of the resonant region) the values for on-site interaction given by the two approaches are al most indistinguishable. This is expected due to the very similar behavior of the bulk profiles of Wannier functions, see Figs. \ref{fig:comparison1} and \ref{fig:comparison3}. 

From Fig.~\ref{fig:integrals} it is also evident that, in the range of $\epsilon$ considered here, the $s-p$ density-density interaction is completely negligible with respect to the on-site interaction. Similarly small is the pair tunneling contribution - as the integrals involved are exactly the same for contact interactions. These findings parallel general discussion for contribution of various interaction terms in the standard optical lattice - for a recent review see \cite{toberopp}. Let us note, following \cite{toberopp}, that the relative role of different terms may be different for long-range, e.g. dipolar, interactions which is, however, beyond the scope of this paper. On the other hand, the significance of the density induced tunnelings 
with respect to the leading order tunneling appearing in $\hat{H}_{0}$ depends on the value of the interaction constant $g$ and the lattice filling factors.  When $g$ is small they can be safely neglected -- see, however, \cite{toberopp} and references therein.

\section{Discussion}

As mentioned in the introduction, superlattice potentials are often proposed to be employed for generating interesting model Hamiltonians. In particular, in this context,  the so called $J_{1}-J_{2}$ (particularly for spins) or $t-t'$ model was invoked
with the non interacting part of the Hamiltonian being of the form 
\begin{equation}
\hat{H}_0=-\sum_{i}\left(t\hat{a}_i^\dag\hat{a}_{i+1}+ t'\hat{a}_i^\dag\hat{a}_{i+2}+h.c. \right)
+\sum_i E_i \hat{n}_i.
\label{eq:final}
\end{equation}
Here $t\equiv J_1$ corresponds to the nearest neighbor hopping  while $t' \equiv J_2$ is the next-nearest neighbor tunneling and $E_i$ stands for the on-site energy, 
common for all sites. By adding different form of interactions, different models
are realized \cite{Dhar2011,dhar13,Li13a} exhibiting frustration and different 
possible phases when arbitrary values of $J_1$ and $J_2$ parameters are assumed. It is often suggested  that these models  may be  realized in one-dimensional superlattices.
The present analysis, together with results of \cite{Modugno} for lowest bands in optical 
superlattices, clearly indicates that it may be quite hard to observe the phases suggested 
in those papers. For lowest bands of superlattices \cite{Modugno} the tunneling
amplitudes are of the same sign while most interesting phases \cite{Dhar2011,dhar13}
apper for opposite signs of $J_1$ and $J_2$. For the discussed here $s-p$ superlattice model,
the tunnelings break left-to-right symmetry so mapping to $J_1-J_2$ model may be realised
by changing the phases of Wannier functions breaking the translational invariance as
discussed at the end of Sec.~\ref{resonance} for the resoant case. The similar approach is possible also out of the
resonance. Still, as discussed extensively above, for the optimal, two-bands approach the nearest neighbour
tunneling always dominates the next nearest terms. Thus it is impossible to reach $J_1\approx 2|J_2|$
region, most interesting for the appearance of Majumdar-Ghosh \cite{majumdar} phase. 
The detailed discussion of possible phases in $s-p$ superlattice in the presence of interactions, 
also including density-dependent tunnelings  deserves a separate investigation.

\acknowledgments
J.Z. acknowledges discussions with M. \L{}\k{a}cki and K. Sacha.
This work has been supported by Polish National Science Centre grant
DEC-2012/04/A/ST2/00088 (WG and JZ),  the UPV/EHU under program UFI 11/55,
the Spanish Ministry of Science and Innovation through Grant No. FIS2012-36673-C03-03,
and the Basque Government through Grant No. IT-472-10.


\begin{thebibliography}{11}
\bibitem{foot1} This construction works perfectly with a symmetric momentum grid that excludes the points $k=-1,0,1$.
\bibitem{foot2} Generally, since there are two choices one may either try both possibilities and see that only one works properly or use e.g. semiclassics to estimate in which sub-well of the superlattice one can expect lower-lying eigenstates.
\end{thebibliography}
\end{document}